# THE SNS RUN PERMIT SYSTEM*

C. Sibley, SNS, Oak Ridge, TN 37831, USA
E. Bjorklund, LANSCE, Los Alamos, NM 87545, USA

## Abstract

The Spallation Neutron Source (SNS) is an accelerator-based neutron source being built in Oak Ridge, Tennessee, by the U.S. Department of Energy. The SNS will provide the most intense pulsed neutron beams in the world for scientific research and industrial development. The facility is scheduled for completion in 2006. The project is a collaboration between Lawrence Berkeley National Lab, Argonne National Lab, Los Alamos National Lab, Oak Ridge National Lab, Brookhaven National Lab, and Jefferson Lab. The Run Permit System (RPS) is a software based system which has several key functions critical for safe operation of the machine. It coordinates machine mode and beam parameter changes with the timing system, verifies the Machine Protection System (MPS) hardware configuration, and provides an interface for masking MPS hardware inputs when necessary. This paper will describe the primary functionality of the Run Permit System and the interface between the Run Permit System, the Machine Protection System, and the timing system.

## 1 INTRODUCTION

The SNS Run Permit system is one level in the overall SNS Machine Protection System (MPS). The purpose of the MPS is to protect *equipment* from being damaged by the beam. It contains one software level (Run Permit), and three hardware levels (Fast Protect Auto-Reset, Fast Protect Latched, and High-QA MPS). More information on the SNS Machine Protection System can be found in [1,2].

The function of the Run Permit portion of the MPS is to 1) set up and verify the machine operating mode, 2) verify that all the relevant equipment is operating and within the desired setpoint ranges for the selected machine mode, 3) verify the equipment masks in the MPS hardware for each machine mode, 4) verify the beam parameters requested are within tolerance for the machine mode, 5) schedule user defined beam pulse parameters, and 6) provide an operator interface to display the status of the MPS and set software masks as appropriate.

## 2 RUN PERMIT MODES

### 2.1 Machine, Beam, and Operating Modes

The SNS "Machine Operating Mode" consists of two parts, a "Machine Mode" and a "Beam Mode." The Machine Mode determines where the beam stops (Linac Dump, Extraction Dump, Target, etc.). The Beam Mode determines the duration and power of the beam. Establishing a Machine Mode requires turning on equipment, verifying beam dumps are ready, etc. before the desired Machine Mode can be made up. Consequently, the Machine Mode will not change very often under normal commissioning, tune-up, and operating conditions. The Beam Mode, on the other hand, can change on a pulse-to-pulse basis. Table 1 lists the Machine and Beam modes for SNS. The Machine mode and maximum beam power or pulse width are selected in the control room via a key switch. The beam mode can be changed to smaller pulse widths from software applications, such as an emittance scan or profile measurement application.

Table 1: SNS Machine and Beam Modes

| Machine Modes | Beam Modes |
|---|---|
| Ion Source | Diagnostics (10 μsec) |
| Diagnostic-Plate | Diagnostics (50 μsec) |
| Linac Dump | Diagnostics (100 μsec) |
| Injection Dump | Full Width (1 msec) |
| Ring | Low Power (7.5 KW) |
| Extraction Dump | Medium Power (200 KW) |
| Target | Full Power (2 MW) |

The Operating Mode for the machine is the combination of Machine and Beam modes. Note that not every Beam Mode can be applied to a given Machine Mode. For example, when the machine is in "Linac Dump" mode, the Beam Mode may not exceed 7.5 KW, and only the "Target" Machine Mode can accommodate the "Full Power" Beam Mode. Table 2 lists the allowable Machine and Beam Mode combinations.

---

* Work supported by U.S. Department of Energy

Table 2: Valid Machine and Beam Mode Combinations

|  | 10 msec | 50 msec | 100 msec | 1 msec | 7.5 KW | 200 KW | 2 MW |
|---|---|---|---|---|---|---|---|
| **Source** | X | X | X | X |  |  |  |
| **D-Plate** | X | X | X | X |  |  |  |
| **L-Dmp** | X | X | X | X | X |  |  |
| **I-Dmp** | X | X | X | X | X | X |  |
| **Ring** | X | X | X |  | X |  |  |
| **E-Dmp** | X | X | X |  | X |  |  |
| **Target** | X | X | X | X | X | X | X |

## 2.2 Diagnostic Modes

The "Diagnostic" Beam Modes exist to limit the integrated current absorbed by intercepting diagnostic devices such as wire-scanners or emittance harps. The pulse width limits are energy dependant. At lower beam energies damage to copper and diagnostic wires is greater so the beam pulse widths are limited to shorter values. This is accomplished by limiting the width of the beam gate.

Before an application program can insert a wire scanner (for example) into the beamline, it must first make a request to the RPS for an operating mode change. The RPS determines the required operating mode based on the device making the request, and the current machine mode. It then interfaces with the timing system to set new limits on the width of the beam gate. Once conditions have been met for the new operating mode, the RPS sets the new mode (this is also done through the timing system) and signals the application program that it may proceed with the wire scan. The operating mode is locked by the RPS until the application signals the wire scans are complete. Wire scanner "Home" signals are inputs into the MPS system. If a wire scan begins before the appropriate diagnostic mode is made up, the MPS will drop beam. Once the wire scan is complete, the RPS returns the operating mode to its previous value.

## 3 RPS USER PULSE SCHEDULING

Various beam pulse profiles can be requested through applications or an EPICS user interface. Each active profile is given a "User_ID" for use in synchronous data acquisition and for feed forward lookup tables in the LLRF systems [3]. The User_ID is broadcast over a Real Time Data Link [4]. There is a limit of 7 active profiles due to memory constraints in the LLRF system. The integrated beam current in the requested profile is checked against pulse width limits for diagnostic machine modes. The integrated power is calculated for the requested repetition rate of the pulse (and other pulses in the request queue) and placed in the queue if the power limitations are within tolerance for the dump in use. Figure 1 shows the operating envelopes for various rep rates and diagnostic modes.

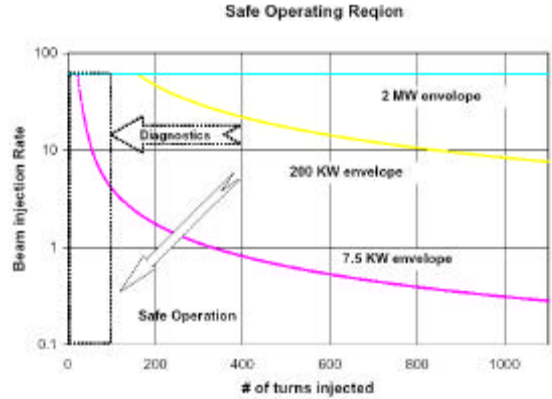

Figure 1. Operating envelopes

## 3.1 Sequencer

The Run Permit System builds the sequences of various pulse types and hands them off to the timing system. Each sequence consists of a "Supercycle" for the machine. This is the lowest continuous repetition rate of interest to operations. Single shot pulses are inserted as requested after pulse width verification. There is a separate sequence for each power level and diagnostic machine mode. The RPS only populates sequences that are allowed in a particular mode and where the mode can change on a pulse-to-pulse basis. For instance, the mode can always change down in power to a diagnostics mode, but can never change to a higher power level than set by operator key settings in the control room. During the operational phase of SNS, diagnostic pulses will be scheduled to calibrate and verify target system diagnostics and the injection phase space painting hardware. A part of the timing system, the Time Line Verification Board monitors the event link and records and verifies the events sent out. Discrepancies result in a temporary beam fault.

## 4 EQUIPMENT MASKING

### 4.1 MPS Mode Masking, Software Masks

The Operating Mode (Machine plus Beam Mode) determines what equipment must be operational and at the correct set points before beam is allowed. The MPS hardware contains a provision for "Mode Masking," which selectively enables or disables inputs into the

MPS system based on the Operating Mode. The mode masks are originally derived from the SNS technical database that describes every piece of equipment and signal in the SNS [5]. The database records whether or not a piece of equipment has an input into the MPS system, and if so, under what Operating Modes the input is relevant. Two files of these masks are created, one for downloading to the IOCs during initialization and a second used by the Run Permit system for verification during running.

Some of the inputs to the MPS can be masked out in software as the need arises. Beam loss monitors for instance, are automatically masked out using software masks when an upstream wire is inserted in the beam. Faulty equipment not required for a particular running scenario can be masked out as long as the MPS hardware configuration allows it. This allows the Mode Masks to remain relatively stable and easily kept in configuration control.

### 4.2 Set Point Monitoring

During the commissioning stages of SNS, the set points of power supplies can be changed as required for various accelerator physics applications, such as beam-based alignment. This should not matter from the machine protection point of view as the average beam power will be small and the accelerator components can generally take two or more full beam pulses in the event of a failure.

In the operational stages the ability to change power supply set points will be restricted. Channel access security will be used to prevent inadvertent set point changes or accidentally powering down a supply. There are other failures that can occur, power supply regulation loops failing, zero flux current transducers failing, or DAC failures in the power supply controller. An EPICS State Notation Language program running on the IOCs will be monitoring the equipment to detect these types of failures. Although there are other systems designed to detect these failures (primarily the Beam Loss Monitor System), this adds to the defense in depth character of the Machine Protection System.

## 5 OPERATOR INTERFACE

The final function of the RPS is to provide an operator interface to the SNS MPS system. Using EPICS RPS screens, the operator can view the state of the MPS system, see which sections are ready to take beam, which sections have MPS faults, and which MPS inputs have been bypassed. A fault timeline is provided to determine where a fault originated. The fault timers use an "Event Link" clock (approximately 16 MHz) to record fault times. These counters are converted to a time stamp in EPICS using revolution frequencies distributed in the Real Time Data Link. In some cases (where allowed by the hardware configuration) an MPS input may be bypassed in software using the RPS operator interface screens. Equipment masking is logged in the EPICS Archiver and acknowledged during shift changes, or as required by operations.

## 6 CONCLUSIONS – CURRENT STATUS

The MPS is designed for Defense in Depth and the RPS is the software layer of that defense. Hardware is used wherever a system failure will cause a loss of beam. Software is used for equipment status verification, including the configuration of the MPS.

The RPS will be installed in phases following the installation of the beamline equipment. The present installation schedule is shown in Table 3.

Table 3. Global Controls (GC) Readiness dates.

| Activity | Finish |
|---|---|
| GC Ready for FE commissioning | 9/20/02 |
| GC Ready for DTL commissioning | 12/4/02 |
| GC Ready for CCL commissioning | 4/27/04 |
| GC Ready for SCL commissioning | 8/19/04 |
| GC Ready for HEBT-Ring comm.. | 11/8/04 |
| GC Ready for RTBT-TGT comm. | 11/15/05 |
| GC Subtask Complete | 11/30/05 |

Although the commissioning and installation schedule shows completion in 2005, the RPS needs basic functionality, pulse scheduling, operator interface, and user defined pulse shapes to be operational for the front end commissioning in 2002.